\documentclass[conference, 10pt]{IEEEtran}
\IEEEoverridecommandlockouts
\usepackage{latexsym}
\usepackage{graphicx}
\usepackage{amsfonts,amssymb,amsmath}

\def\nbc{{\mathbf{c}}}

\def\nbu{{\mathbf{u}}}
\def\nbv{{\mathbf{v}}}

\def\nbx{{\mathbf{x}}}

\def\nbz{{\mathbf{z}}}
\def\nb0{{\mathbf{0}}}
\def\nb1{{\mathbf{1}}}

\def\nbC{{\mathbf{C}}}

\def\nbG{{\mathbf{G}}}

\def\N{\sigma^2}

\usepackage[acronyms,nonumberlist,nopostdot,nomain,nogroupskip]{glossaries}
\newacronym{quic}{QUIC}{Quick UDP Internet Connections}
\newacronym{3gpp}{3GPP}{3rd Generation Partnership Project}
\newacronym{adc}{ADC}{Analog to Digital Converter}
\newacronym{5g}{5G}{5th generation}
\newacronym{aimd}{AIMD}{Additive Increase Multiplicative Decrease}
\newacronym{am}{AM}{Acknowledged Mode}
\newacronym{amc}{AMC}{Adaptive Modulation and Coding}
\newacronym{aqm}{AQM}{Active Queue Management}
\newacronym{awgn}{AGWN}{Additive White Gaussian Noise}
\newacronym{afd}{AFD}{Austin Fire Department}
\newacronym{balia}{BALIA}{Balanced Link Adaptation}
\newacronym{bdp}{BDP}{Bandwidth-Delay Product}
\newacronym{bf}{BF}{Beamforming}
\newacronym{cc}{CC}{Congestion Control}
\newacronym{cdf}{CDF}{Cumulative Distribution Function}
\newacronym{cn}{CN}{Core Network}
\newacronym{cqi}{CQI}{Channel Quality Information}
\newacronym{cp}{CP}{Control Plane}
\newacronym{csirs}{CSI-RS}{Channel State Information - Reference Signal}
\newacronym{dc}{DC}{Dual Connectivity}
\newacronym{dce}{DCE}{Direct Code Execution}
\newacronym{dci}{DCI}{Downlink Control Information}
\newacronym{dl}{DL}{Downlink}
\newacronym{dmr}{DMR}{Deadline Miss Ratio}
\newacronym{dmrs}{DMRS}{DeModulation Reference Signal}
\newacronym{e2e}{E2E}{End-to-End}
\newacronym{ecn}{ECN}{Explicit Congestion Notification}
\newacronym{edf}{EDF}{Earliest Deadline First}
\newacronym{enb}{eNB}{evolved Node Base}
\newacronym{epc}{EPC}{Evolved Packet Core}
\newacronym{es}{ES}{Edge Server}
\newacronym{fdma}{FDMA}{Frequency Division Multiple Access}
\newacronym{fdd}{FDD}{Frequency Division Duplexing}
\newacronym[firstplural=Radio Access Technologies (RATs)]{rat}{RAT}{Radio Access Technology}
\newacronym{fs}{FS}{Fast Switching}
\newacronym{ftp}{FTP}{File Transfer Protocol}
\newacronym{gnb}{gNB}{Next Generation Node Base}
\newacronym{harq}{HARQ}{Hybrid Automatic Repeat reQuest}
\newacronym{hetnet}{HetNet}{Heterogeneous Network}
\newacronym{hh}{HH}{Hard Handover}
\newacronym{hol}{HOL}{Head-of-Line}
\newacronym{ia}{IA}{Initial Access}
\newacronym{imt}{IMT}{International Mobile Telecommunication}
\newacronym{iot}{IoT}{Internet of Things}
\newacronym{los}{LOS}{Line of Sight}
\newacronym{lte}{LTE}{Long Term Evolution}
\newacronym{m2m}{M2M}{Machine to Machine}
\newacronym{mac}{MAC}{Medium Access Control}
\newacronym{mc}{MC}{Multi-Connectivity}
\newacronym{mcs}{MCS}{Modulation and Coding Scheme}
\newacronym{mec}{MEC}{Mobile Edge Cloud}
\newacronym{mi}{MI}{Mutual Information}
\newacronym{mimo}{MIMO}{Multiple Input, Multiple Output}
\newacronym{mmwave}{mmWave}{millimeter wave}
\newacronym{mr}{MR}{Maximum Rate}
\newacronym{mss}{MSS}{Maximum Segment Size}
\newacronym{mtd}{MTD}{Machine-Type Device}
\newacronym{mtu}{MTU}{Maximum Transmission Unit}
\newacronym{nfv}{NFV}{Network Function Virtualization}
\newacronym{nlos}{NLOS}{Non Line of Sight}
\newacronym{nr}{NR}{New Radio}
\newacronym{ofdm}{OFDM}{Orthogonal Frequency Division Multiplexing}
\newacronym{pdcch}{PDCCH}{Physical Downlonk Control Channel}
\newacronym{pdcp}{PDCP}{Packet Data Convergence Protocol}
\newacronym{pdsch}{PDSCH}{Physical Downlink Shared Channel}
\newacronym{pdu}{PDU}{Packet Data Unit}
\newacronym{pf}{PF}{Proportional Fair}
\newacronym{pgw}{PGW}{Packet Gateway}
\newacronym{phy}{PHY}{Physical}
\newacronym{pbch}{PBCH}{Physical Broadcast Channel}
\newacronym[plural=\gls{mme}s,firstplural=Mobility Management Entities (MMEs)]{mme}{MME}{Mobility Management Entity}
\newacronym{prb}{PRB}{Physical Resource Block}
\newacronym{pss}{PSS}{Primary Synchronization Signal}
\newacronym{pucch}{PUCCH}{Physical Uplink Control Channel}
\newacronym{pusch}{PUSCH}{Physical Uplink Shared Channel}
\newacronym{rach}{RACH}{Random Access Channel}
\newacronym{ran}{RAN}{Radio Access Network}
\newacronym{red}{RED}{Robotics Emergency Deployment}
\newacronym{rf}{RF}{Radio Frequency}
\newacronym{rlc}{RLC}{Radio Link Control}
\newacronym{rlf}{RLF}{Radio Link Failure}
\newacronym{rrc}{RRC}{Radio Resource Control}
\newacronym{rrm}{RRM}{Radio Resource Management}
\newacronym{rr}{RR}{Round Robin}
\newacronym{rs}{RS}{Remote Server}
\newacronym{rsrp}{RSRP}{Reference Signal Received Power}
\newacronym{rss}{RSS}{Received Signal Strength}
\newacronym{rtt}{RTT}{Round Trip Time}
\newacronym{rw}{RW}{Receive Window}
\newacronym{rx}{RX}{Receiver}
\newacronym{sa}{SA}{standalone}
\newacronym{sack}{SACK}{Selective Acknowledgment}
\newacronym{sap}{SAP}{Service Access Point}
\newacronym{sch}{SCH}{Secondary Cell Handover}
\newacronym{scoot}{SCOOT}{Split Cycle Offset Optimization Technique}
\newacronym{sdma}{SDMA}{Spatial Division Multiple Access}
\newacronym{sinr}{SINR}{Signal to Interference plus Noise Ratio}
\newacronym{sm}{SM}{Saturation Mode}
\newacronym{snr}{SNR}{Signal to Noise Ratio}
\newacronym{son}{SON}{Self-Organizing Network}
\newacronym{ss}{SS}{Synchronization Signal}
\newacronym{srs}{SRS}{Sounding Reference Signal}
\newacronym{sss}{SSS}{Secondary Synchronization Signal}
\newacronym{tb}{TB}{Transport Block}
\newacronym{tcp}{TCP}{Transmission Control Protocol}
\newacronym{tdd}{TDD}{Time Division Duplexing}
\newacronym{tdma}{TDMA}{Time Division Multiple Access}
\newacronym{tfl}{TfL}{Transport for London}
\newacronym{tm}{TM}{Transparent Mode}
\newacronym{trp}{TRP}{Transmitter Receiver Pair}
\newacronym{tti}{TTI}{Transmission Time Interval}
\newacronym{ttt}{TTT}{Time-to-Trigger}
\newacronym{tx}{TX}{Transmitter}
\newacronym{ue}{UE}{User Equipment}
\newacronym{ul}{UL}{Uplink}
\newacronym{uml}{UML}{Unified Modeling Language}
\newacronym{um}{UM}{Unacknowledged Mode}
\newacronym{utc}{UTC}{Urban Traffic Control}
\newacronym{vm}{VM}{Virtual Machine}
\newacronym{rsrq}{RSRQ}{Reference Signal Received Quality}
\newacronym{rssi}{RSSI}{Received Signal Strength Indicator}
\newacronym{crs}{CRS}{Cell Reference Signal}
\newacronym{comp}{CoMP}{Coordinated Multi-Point}
\newacronym{cran}{C-RAN}{Cloud \acrlong{ran}}
\newacronym{ca}{CA}{Carrier Aggregation}
\newacronym{cco}{CC}{Carrier Component}
\newacronym{nsa}{NSA}{Non Stand Alone}
\newacronym{embb}{eMBB}{Enhanced Mobility Broadband}
\newacronym{bsr}{BSR}{Buffer Status Report}
\newacronym{srb}{SRB}{Service Radio Bearer}
\newacronym{scm}{SCM}{Spatial Channel Model}
\newacronym{sctp}{SCTP}{Stream Control Transmission Protocol}
\newacronym{mptcp}{MPTCP}{Multi-path TCP}
\newacronym{ietf}{IETF}{Internet Engineering Task Force}
\newacronym{os}{OS}{Operating System}
\newacronym{tls}{TLS}{Transport Layer Security}
\newacronym{rfc}{RFC}{Request for Comments}
\newacronym{http}{HTTP}{HyperText Transfer Protocol}
\newacronym{nat}{NAT}{Network Address Translation}
\newacronym{api}{API}{Application Programming Interface}
\newacronym{rto}{RTO}{Retransmission Timeout}
\newacronym{psc}{PSC}{Public Safety Communication}
\newacronym{rpgm}{RPGM}{Reference Point Group Mobility}
\newacronym{ic}{IC}{Incident Command}
\newacronym{rsu}{RSU}{Road Side Unit}
\newacronym{uav}{UAV}{unmanned aerial vehicle}
\newacronym{usv}{USV}{Unmanned Surface Vehicle}
\newacronym{uas}{UAS}{Unmanned Aerial System}
\newacronym{iab}{IAB}{Integrated Access and Backhaul}
\newacronym{qoe}{QoE}{Quality of Experience}
\newacronym{ssim}{SSIM}{Structural Similarity Index}
\newacronym{psnr}{PSNR}{Peak Signal to Noise Ratio}
\newacronym{bs}{BS}{Base Station}
\newacronym{mu}{MU}{Multiple User}
\newacronym{ag}{AG}{Air-to-Ground}
\newacronym{af}{AF}{Array Factor}
\newacronym{ula}{ULA}{Uniform Linear Array}
\newacronym{upa}{UPA}{Uniform Planar Array}
\newacronym{lcs}{LCS}{Local Coordinate System}
\newacronym{psd}{PSD}{Power Spectral Density}
\newacronym{vq}{VQ}{vector quantization}
\newacronym{a2g}{A2G}{air-to-ground}
\newacronym{em}{EM}{electromagnetic}
\newacronym{vae}{VAE}{variational autoencoder}

\def\bb0{{\mathbb{0}}}

\def\bb{{\boldsymbol{b}}}

\def\b0{{\boldsymbol{0}}}

\def\b{{\mathrm{b}}}

\def\r0{{\mathbf{0}}}

\def\bsf0{{\bm{\mathsf{0}}}}

\def\N0{{N_{\mathrm{0}}}}

\def\bsf{{\boldsymbol{s}_\mathrm{f}}}

\newcommand{\be}{\begin{equation}}
\newcommand{\ee}{\end{equation}}
\newcommand{\bal}{\begin{align}}
\newcommand{\eal}{\end{align}}

\usepackage{booktabs} 
\usepackage{tabularx} 
\usepackage{textcomp}
\usepackage{gensymb}
\usepackage{siunitx}
\usepackage{xcolor}
\usepackage{tikz}
\usepackage{etoolbox}
\usepackage{enumitem}
\usepackage{esvect}
\usepackage{mathtools}
\newtoggle{conf}
\toggletrue{conf}
\usetikzlibrary{chains,arrows,calc,positioning}
\usepackage{multirow}
\usepackage{comment}
\usepackage[caption=false, font=footnotesize]{subfig}
\usepackage[normalem]{ulem}

\newtoggle{conference}
\toggletrue{conference}  

\def\BibTeX{{\rm B\kern-.05em{\sc i\kern-.025em b}\kern-.08em T\kern-.1667em\lower.7ex\hbox{E}\kern-.125emX}}

\begin{document}
%
\title{Channel Modeling for FR3 Upper Mid-band \\
via Generative Adversarial Networks}
\author{
\IEEEauthorblockN{Yaqi Hu\IEEEauthorrefmark{1}, Mingsheng Yin\IEEEauthorrefmark{1}, Marco Mezzavilla\IEEEauthorrefmark{2}, Hao Guo\IEEEauthorrefmark{1}\IEEEauthorrefmark{3}, Sundeep Rangan\IEEEauthorrefmark{1}} 
\IEEEauthorblockA{\IEEEauthorrefmark{1}NYU Tandon School of Engineering, Brooklyn, NY, USA\\
\IEEEauthorrefmark{2}Dipartimento di Elettronica, Informazione e Bioingegneria (DEIB), Politecnico di Milano, Milan, Italy\\
\IEEEauthorrefmark{3}Department of Electrical Engineering, Chalmers University of Technology, Gothenburg, Sweden\\
}

\thanks{The work was supported, in part, by NSF grants 1952180, 2133662, 2236097, 2148293, 1925079, the NTIA, the industrial affiliates of NYU WIRELESS, and by Remcom that provided the Wireless InSite  
software. Hao Guo was supported by the Swedish Research Council (No. 2023-00272).}
}
\IEEEaftertitletext{\vspace{-2\baselineskip}}
\maketitle

\begin{abstract}
The upper mid-band (FR3) has been recently attracting interest for new generation of mobile networks, as it provides a promising balance between spectrum availability and coverage, which are inherent limitations of the sub 6GHz and millimeter wave bands, respectively. In order to efficiently design and optimize the network, channel modeling plays a key role since FR3 systems are expected to operate at multiple frequency bands. Data-driven methods, especially generative adversarial networks (GANs), can capture the intricate relationships among data samples, and provide an appropriate tool for FR3 channel modeling. In this work, we present the architecture, link state model, and path generative network of GAN-based FR3 channel modeling. The comparison of our model greatly matches the ray-tracing simulated data.
\end{abstract}


\IEEEpeerreviewmaketitle

\section{Introduction}
\label{sec:intro}
With congested sub-\SI{6}{GHz} and the blockage/short-range issues at millimeter wave bands, FR3, also referred to as the upper mid-band (from approximately \SI{7}{GHz} to \SI{24}{GHz}) has been attracting research interests from both academic and industrial perspectives \cite{kang2024cellular}. FR3 provides an excellent balance of coverage and bandwidth, while the range of spectrum is large on its own. Nevertheless, with current incumbents such as military radar, radio astronomy, and communication satellites, the radio spectrum resources at FR3 will likely need to be shared. 

Spectrum sharing in the upper mid-band has been considered a key challenge in order to fully exploit the FR3 potential, and it will call for channel modeling efforts spanning multiple carrier frequencies. Here, the desired models should be able to capture the joint distribution across multiple bands. There have been recent multi-frequency measurement campaigns \cite{huang2020multi,cui2020multi}. However, these measurements are enormously time-consuming, and adaptive models are still missing.

Data-driven machine-learning methods with minimal assumptions can naturally capture intricate probabilistic relationships. Generative neural networks provide a natural approach
to data-driven channel modeling that can broadly represent complex settings, and some early works have successfully trialed generative adversarial networks (GANs) for simple wireless channels \cite{gulrajani2017improved,arjovsky2017wasserstein}. In our previous study \cite{XiaRanMez2020}, we propose a generative model that is constructed from two cascaded neural networks as shown in Fig.~\ref{fig:gen_model}: The first network generates the link state (i.e., if the link is in line-of-sight, non-line of sight, or outage) based on the condition vector $\nbu$. The second network generates the path data vector based on the condition vector and link state. Nevertheless, the method proposed in \cite{XiaRanMez2020} cannot be applied directly to FR3 due to its multi-frequency properties. Moreover, the channel measurements \cite{huang2020multi,cui2020multi} do not provide sufficient data points for training complex neural networks. 

In this paper, we propose a GAN-based channel modeling method that aims to capture  the characteristics of FR3 multi-band operations. The key differences from \cite{XiaRanMez2020} are 1) the path generator network uses a GAN, and 2) the data is generated across multiple frequencies. The contributions of the paper are summarized as follows:

\begin{itemize}
    \item We consider the use of the multi-frequency GAN channel methodology developed in \cite{hu2022multi}.
    The benefit of this procedure is that the cross-frequency relations are well captured.   The work \cite{hu2022multi} leveraged the framework for mmWave frequencies.  Here, we apply the framework to the upper mid-band and improve the adapt the model in \cite{hu2022multi} to upper-mid band by a link state predictor.
    \item We demonstrate that the GAN modeling framework can capture key aspects of 
    the wideband channel. For instance, it can capture the marginal and \emph{joint} distributions of the path loss.
    \item We demonstrate how the method can be used for assessing inter-frequency tasks.  In particular, we evaluate a beamforming approach where the beamforming vector is selected at a lower frequency
    and then applied at a higher frequency.
    \item The proposed method is also evaluated by the angles spread of both AoA and AoD for different frequencies.
\end{itemize}

\section{System Model}

As an extension of our previous study \cite{hu2022multi} where we presented a general methodology for modeling full double-directional channels at multiple $M$ frequencies, here, we focus on FR3/upper mid-band \cite{kang2024cellular} and model channel with $M = 4$ at $f_1=$\, \SI{6}{GHz}, $f_2=$\, \SI{12}{GHz}, $f_3=$\, \SI{18}{GHz}, and $f_4=$\ \SI{24}{GHz}. In this section, we first present the multi-frequency channel model specifically for FR3. Then, our proposed generative neural network-based modeling method is introduced with a two-stage structure.

\subsection{Multi-Frequency Channel Model for FR3}
\label{sec:model}




\begin{figure*}[ht]
\footnotesize
\centering
\includegraphics[width=0.8 \linewidth]{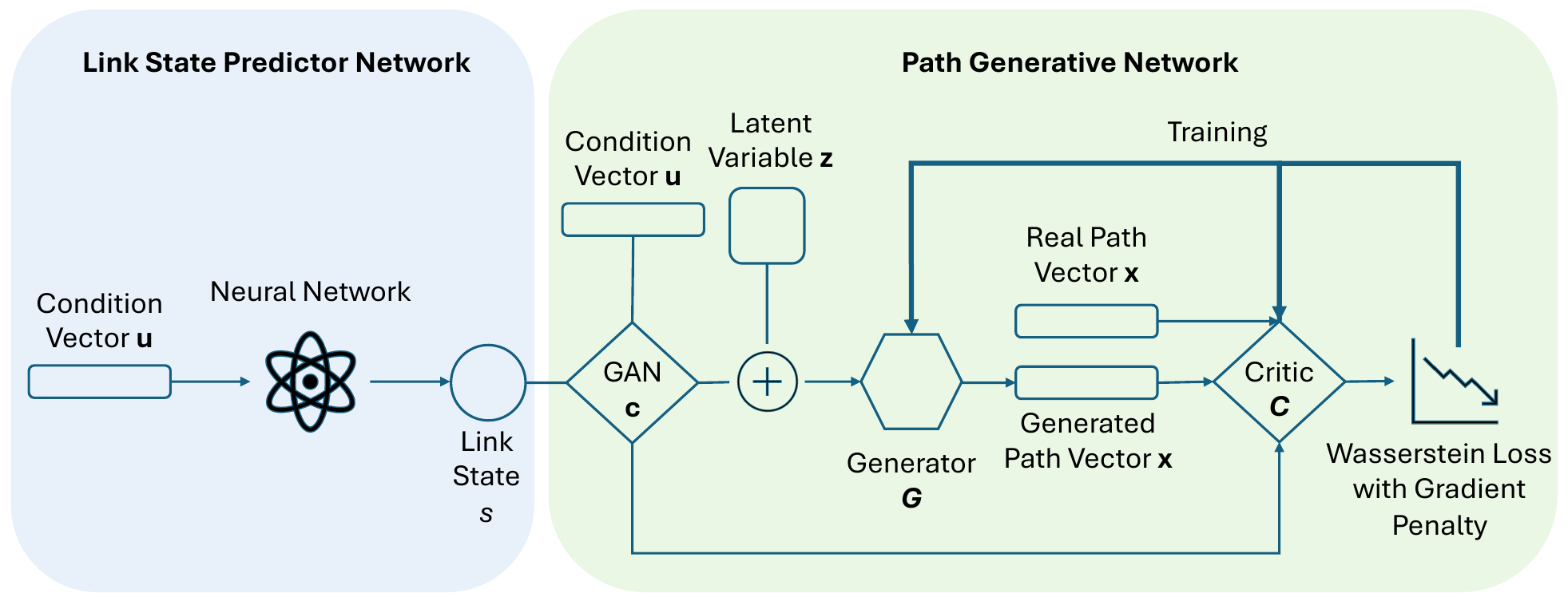}
\caption{The architecture for the proposed two-stage generative neural network model.}
\label{fig:gen_model}
\vspace{-4mm}
\end{figure*}


In this work, we model a single link, i.e.,  the channel from a transmitter (TX) to receiver (RX). For instance, the channel between a base station (gNB) and a user equipment (UE) as in cellular applications. With reciprocity, the TX and RX could be reversed. We assume that the channel across the $M$ frequencies share a common set of $L$ paths where each path is described by a vector of parameters \cite{hu2022multi}, 
\begin{equation} \label{eq:thetaell}
    \chi_\ell
        = \left[\tau_\ell, 
        \theta^{\rm rx}_\ell, \phi^{\rm rx}_\ell, 
        \theta^{\rm tx}_\ell, \phi^{\rm tx}_\ell, 
        G_{\ell 1},\ldots,G_{\ell M} \right],
\end{equation}
where $\tau_\ell$ is the path delay, $\theta^{\rm rx}_\ell$, $\phi^{\rm rx}_\ell$ are the elevation and azimuth angles of arrival (AoA), $\theta^{\rm tx}_\ell$, $\phi^{\rm tx}_\ell$ are the elevation and azimuth angles of departure (AoD), and $G_{\ell 1},\ldots,G_{\ell M}$ are the path gains across the $M$ frequencies. Here, the implicit assumption is that the path angles and delays are shared across the frequencies. It has been used in the 3GPP statistical model \cite{3GPP38901} and is physically justified since the waves between the TX and RX at different frequencies generally propagate over the same paths.
The multi-frequency channel can then be described by the set of $L$ paths \begin{equation} \label{eq:theta}
    \nbx = \left[ \chi_1, \ldots, \chi_L \right].
\end{equation}
We will call $\nbx$ the \emph{path vector}. This channel description is similar to \cite{XiaRanMez2020} but with $M$ path gains per path due to FR3 characteristics. Also, we set $L=20$ as it is the default maximum number of paths found by the ray tracer as described below. When there are less than $L$ paths, say $L_0 < L$, we simply set $G_{\ell m}=0$ (in linear scale) for all $\ell > L_0$ and all $m=1,\ldots,M$. Moreover, each link also has a condition vector, $\nbu$, which is defined as
\begin{equation} \label{eq:uvec} 
    \nbu = \left(d_x,d_y,d_z\right), 
\end{equation} 
i.e., the 3D vector between the TX and RX. Other parameters could be added to $\nbu$ such as the antenna heights or cell types. 


The statistical channel modeling problem is to model the distribution of the path vector $\nbx$ as a function of the link conditions $\nbu$, i.e., the conditional probability distribution $p(\nbx|\nbu)$. In this work, we deploy the so-called generative models and the conditional distribution we are looking for can be represented as a mapping \begin{equation} \label{eq:gen} 
    \nbx = g(\nbu, \nbz), 
\end{equation} 
where $\nbz$ is some random vector with a known distribution. As indicated in \cite{3GPP38901}, generative models provide powerful tools for the evaluations of cellular networks, especially for the multiple-input multiple-output (MIMO) channels. The randomness of the radio channel can be well captured by generative models with path parameters and antenna array assumptions.


\subsection{Generative Neural Network Modeling}
As shown in Fig. \ref{fig:gen_model}, our proposed generative neural network model contains two stages: link state predictor network and path generative network.
\subsubsection{Link State Model}
The first network -- the link state network -- takes as an input the condition vector $\nbu$
in \eqref{eq:uvec} and
generates a random link state, denoted as $s$, corresponding to one of three states  \cite{akdeniz2014millimeter}:
\begin{enumerate}[label=\roman*)]
    \item \verb|LOS|: The LOS path is present, in addition to NLOS paths;
    \item \verb|NLOS|: No LOS path, but at least one NLOS path is present;
    \item \verb|Outage|: No propagation paths (either LOS or NLOS) exist for this link.
\end{enumerate}
The structure of this network is based on \cite{XiaRanMez2020}.
Specifically, the condition vector $\nbu$ is first converted to a 3D distance and 
elevation distance, denoted as $\nbu=(d_\text{3D}, d_z)$.  Then, the vector $\nbu$
is fed to a fully connected neural network with parameters shown in 
Table~\ref{table:model_config}.  The output of the network is a 3-way softmax that 
generates the probabilities of each link state.

\subsubsection{Path Generative Network}  The input to the path generative network
is the GAN vector $\nbc = (\nbu, s)$ which contains the 
link condition vector $\nbu$ along with the randomly generated link state $s$.
The goal of the network is to then generate random path vectors $\nbx$ in
\eqref{eq:theta} following the conditional distribution $p(\nbx|\nbc)$
as observed in the data.  

For this purpose, we use a 
conditional Wasserstein generative adversarial network with gradient penalty (CWGAN-GP) \cite{gulrajani2017improved,arjovsky2017wasserstein} with two components:  
a \emph{generator} $\nbG$
and \emph{critic} $\nbC$.  The generator takes the condition vector $\nbc$ and some
random input $\nbz \sim p(\nbz)$ and generates a random path data vector $\nbx = \nbG(\nbc,\nbz)$ where the vector $\nbz$ is called the \emph{latent vector}.
As is commonly used in \cite{gulrajani2017improved,arjovsky2017wasserstein}, we take its distribution $p(\nbz)$ to be a unit variance Gaussian vector.  Moreover, we set the latent dimension to 20.
To ensure the generator matches the data distribution, one can also train a critic
function $\nbC$ that attempts to discriminate between the generated and true samples.
Finally, the generated network is trained with a loss function defined as
\begin{align}
    \label{eq:wasser_loss}\underbrace{\underset{\tilde{\boldsymbol{x}} \sim \mathbb{P}_{g}}{\mathbb{E}}\left[C(\tilde{\boldsymbol{x}})\right]-\underset{\boldsymbol{x} \sim \mathbb{P}_{r}}{\mathbb{E}}\left[C(\boldsymbol{x})\right]}_{\text {Original critic loss }}+ \nonumber 
    \underbrace{\lambda \underset{\hat{\boldsymbol{x}} \sim \mathbb{P}_{\hat{\boldsymbol{x}}}}{\mathbb{E}}\left[\left(\left\|\nabla_{\hat{\boldsymbol{x}}} D(\hat{\boldsymbol{x}})\right\|_{2}-1\right)^{2}\right] ,}_{\text {Gradient penalty }}
\end{align} 
where  $\mathbb{P}_g$ is the generator distribution implicitly defined by $\tilde{\boldsymbol{x}}=\nbG(\boldsymbol{c},\boldsymbol{z})$, with $\boldsymbol{z} \sim p(\boldsymbol{z})$. Also, $\mathbb{P}_r$ is the distribution of the data points.  The loss
is optimized via a minimax operation:  The critic $\nbC$ attempts to minimize the loss
to discriminate between the generated and true samples, while the 
generators attempt to maximize the loss to fake the critic.  The minimax is similar to other GANs,
but the key concept in the WGAN-GP is that the gradient penalty term avoids mode collapse
-- see \cite{gulrajani2017improved,arjovsky2017wasserstein} for details.  
For our application, both the generator and critic are realized with as fully connected
neural networks with parameters shown in Table \ref{table:model_config}. Our implementation details are available online at~\cite{mmwchanmod-GAN}.


\begin{table}[t]
\caption{Generative model configuration}
\begin{center}
\begin{tabular}{l|c|c|c|}
\cline{2-4}
                                          & \begin{tabular}[c]{@{}c@{}}Link state \\ prediction\end{tabular} & \begin{tabular}[c]{@{}c@{}}Path GAN \\ critic\end{tabular} & \begin{tabular}[c]{@{}c@{}}Path GAN \\ generator\end{tabular} \\ \hline
\multicolumn{1}{|l|}{Num. of inputs}     & 2                                                                & 180 $+$ 3                                                & 20 $+$ 3                                                      \\ \hline
\multicolumn{1}{|l|}{Hidden units}         & {[}25, 10{]}                                                     & {[}1120,560,280{]}                                         & {[}280,560,1120{]}                                            \\ \hline
\multicolumn{1}{|l|}{Num. of outputs}    & 3                                                                & 1                                                          & 180                                                           \\ \hline
\multicolumn{1}{|l|}{Optimizer}            & Adam                                                             & \multicolumn{2}{c|}{Adam}                                                                                                  \\ \hline
\multicolumn{1}{|l|}{Learning rate}        & 0.001                                                            & \multicolumn{2}{c|}{0.0001}                                                                                                \\ \hline
\multicolumn{1}{|l|}{Epochs}               & 100                                                               & \multicolumn{2}{c|}{30000}                                                                                                  \\ \hline
\multicolumn{1}{|l|}{Batch size}           & 200                                                              & \multicolumn{2}{c|}{2048}                                                                                                  \\ \hline
\end{tabular} 

\end{center}
\label{table:model_config}
\vspace{-0.5cm}
\end{table}
\section{Data Acquisition: Urban FR3 Cellular System with Ray Tracing}
\label{sec:raytracing}

\begin{figure}
  \centering
  \includegraphics[width=0.9\linewidth]{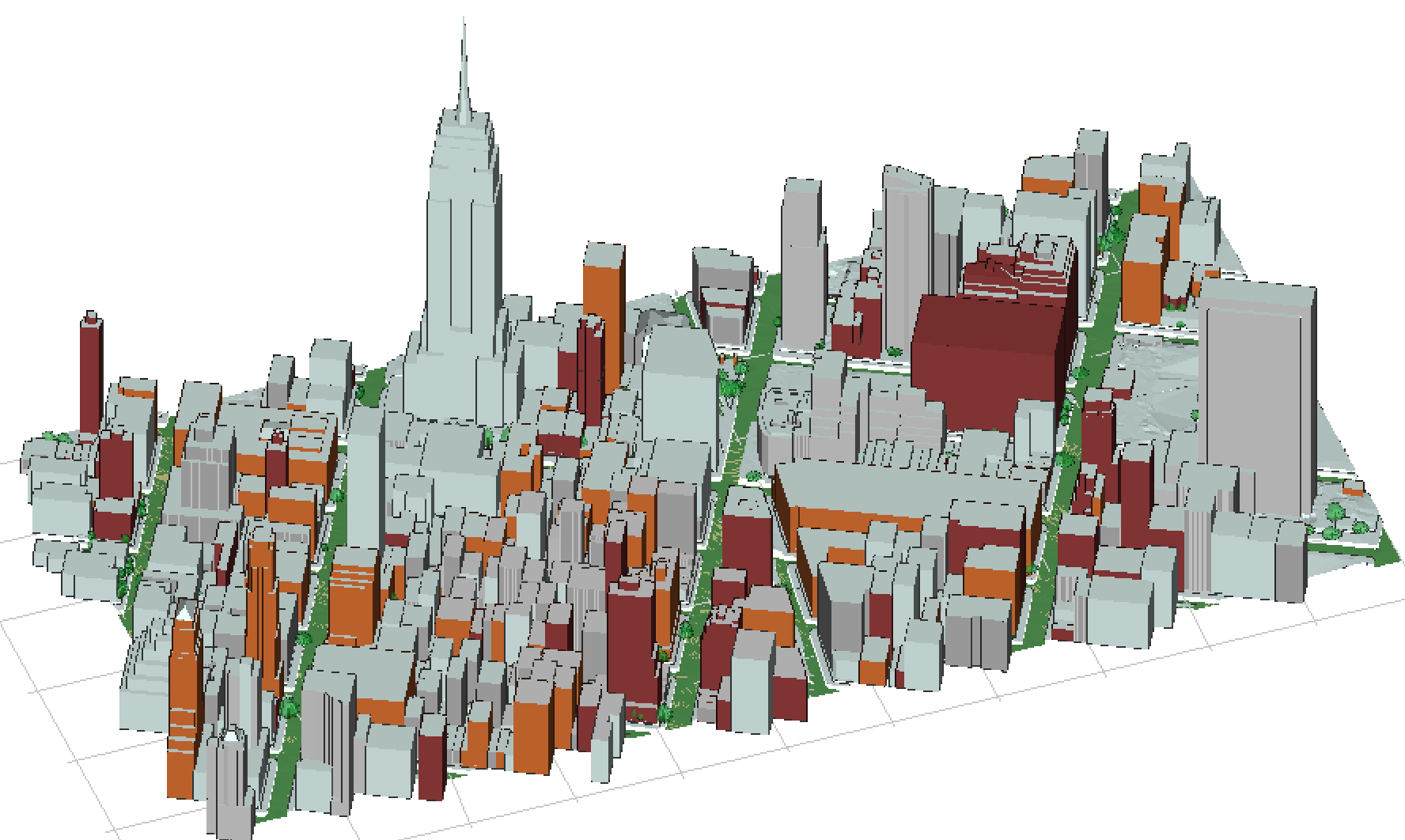}
  \caption{Ray tracing simulation area representing a 1100 $\times$ 570 $\text{m}^2$ urban region of NYC.}
  \label{fig:ray_tracing_map}
  \vspace{-0.7cm}
\end{figure}


This methodology is broadly applicable, yet for this study, we focus on evaluating hypothetical systems at four frequencies of \SI{6}{}, \SI{12}{}, \SI{18}{}, and \SI{24}{GHz} in the upper mid-band band. 
In alignment with recent studies \cite{hu2023parametrization,yin2022millimeter,khawaja2017uav}, we employ an advanced ray tracing tool, Wireless InSite by Remcom \cite{Remcom}, for generating the required datasets for our analysis.


In this work, similar to the configuration in \cite{kang2024cellular}, we target a specific area of New York City covering around 1100 $\times$ 570 square meters, as shown in Fig.~\ref{fig:ray_tracing_map}.
We strategically placed 79 gNBs on building rooftops, achieving a 3-dimensional inter-site distance (ISD) ranging from 100 to 200 meters. This spacing is typical for urban microcellular networks.
Additionally, terrestrial UEs were methodically placed 5 meters apart in a grid layout. 
Importantly, all UEs are positioned \SI{1.6}{m} above the ground level, and some are located inside buildings to comprehensively simulate the urban cellular landscape.
This setup comprises 22,950 UEs, thus establishing 79 $\times$ 22950 $=$ 1,813,050 gNB-UE links.

Our ray tracing simulations were conducted independently at the aforementioned four frequencies. 
We configured the simulations to consider up to 6 reflections, 1 diffraction, and 1 transmission per path, with no more than 20 paths per link. 
Moreover, this ray tracing tool captures only paths with a minimum power level of -250 dBm.

\begin{table}[t]
\caption{Material Properties at Different Frequencies: Conductivity (\SI{}{S/m}) and Permittivity (Relative) \cite{ITUR_P2040_2}}
\label{table:material_properties}
\small 
\setlength{\tabcolsep}{3pt} 
\renewcommand{\arraystretch}{1.1} 
\begin{tabularx}{\columnwidth}{@{}Xcccccccc@{}}
\toprule
Material & \multicolumn{2}{c}{6 GHz} & \multicolumn{2}{c}{12 GHz} & \multicolumn{2}{c}{18 GHz} & \multicolumn{2}{c}{24 GHz} \\
\cmidrule(lr){2-3} \cmidrule(lr){4-5} \cmidrule(lr){6-7} \cmidrule(l){8-9}
         & Con. & Per. & Con. & Per. & Con. & Per. & Con. & Per. \\
\midrule
Concrete & .188 & 5.24 & .323 & 5.24 & .443 & 5.24 & .555 & 5.24 \\
Brick    & .032 & 3.91 & .035 & 3.91 & .038 & 3.91 & .040 & 3.91 \\
Glass    & .040 & 6.31 & .100 & 6.31 & .173 & 6.31 & .254 & 6.31 \\
Plaster  & .046 & 2.73 & .088 & 2.73 & .128 & 2.73 & .168 & 2.73 \\
Wood     & .032 & 1.99 & .067 & 1.99 & .104 & 1.99 & .142 & 1.99 \\
Dry Earth Ground& .649 &12.5 & 2.00 & 10.0 & 4.00 & 7.20 & 5.60 & 6.00 \\
\bottomrule
\end{tabularx}
\vspace{-0.5cm}
\end{table}

The model depicted in Fig.~\ref{fig:ray_tracing_map} is derived from \cite{geoPipe}, which provides an accurate depiction of the characteristics of both foliage and building materials. 
To ensure precise ray tracing across the different frequencies, the properties of materials such as concrete, brick, plaster, wood, glass, and dry earth ground were meticulously detailed, as shown in Table.~\ref{table:material_properties}.


\section{Modeling Results}
\label{sec:modeling_results}

In this section, we present numerical results on the effectiveness of the proposed GAN-based method applied at FR3. As introduced in Section~\ref{sec:raytracing}, the dataset includes 1,813,050 links, each consisting of paths at \SI{6}{GHz}, \SI{12}{GHz}, \SI{18}{GHz}, and \SI{24}{GHz}, respectively. For our analysis, we randomly selected 30\% of the links for training and reserved 10\% of the remaining links for testing.

\subsection{Path Loss with Omnidirectional Antennas}
\begin{figure}
  \centering
  \includegraphics[width=0.88 \linewidth]{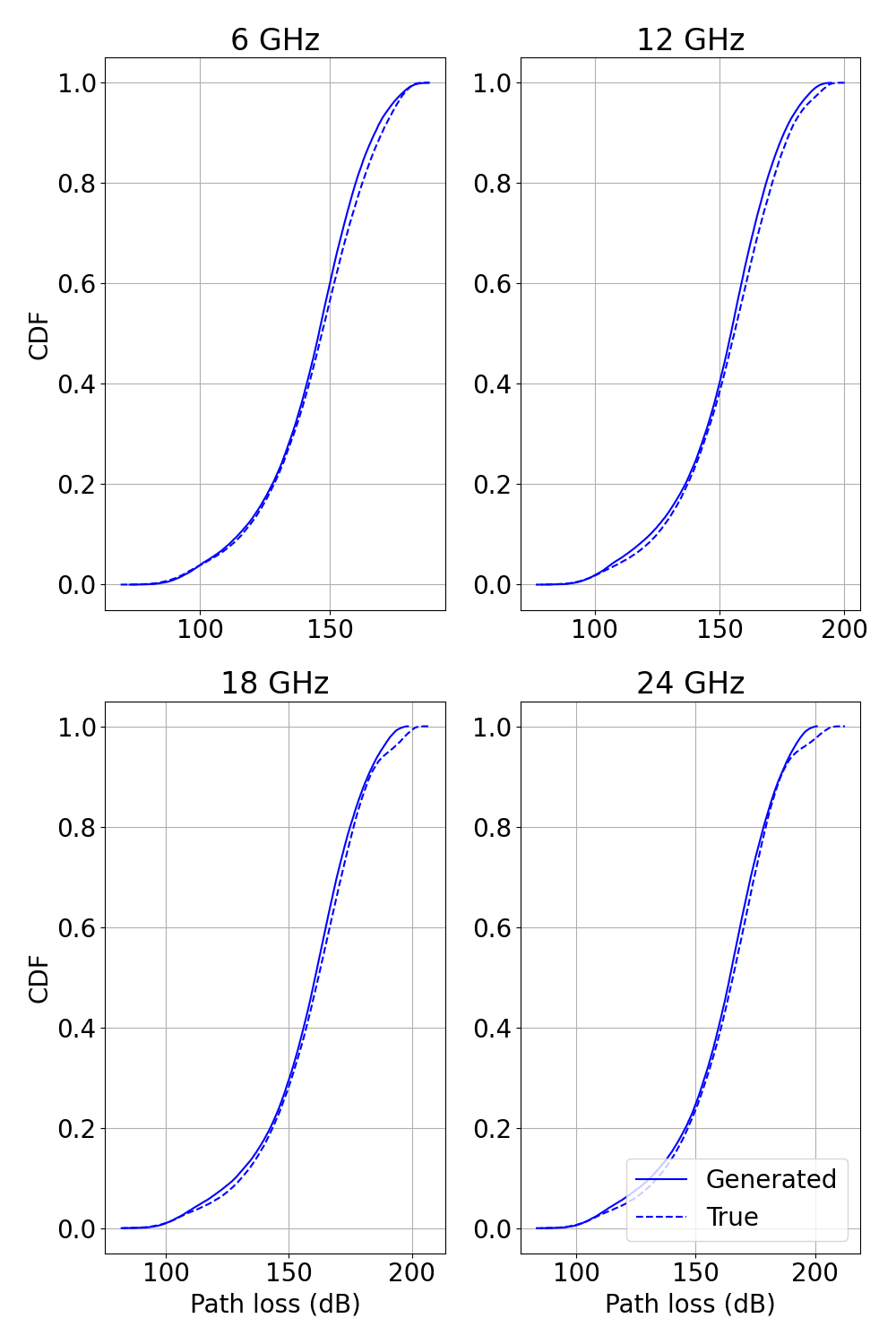}
  \caption{The CDF of the path loss for links within the test dataset is compared against the CDF of the pass loss for links randomly generated by the trained model.}
  \label{fig:received_power_cdf}
  \vspace{-0.7cm}
\end{figure}

\begin{figure*}
  \centering
  \includegraphics[width=1.0 \linewidth]{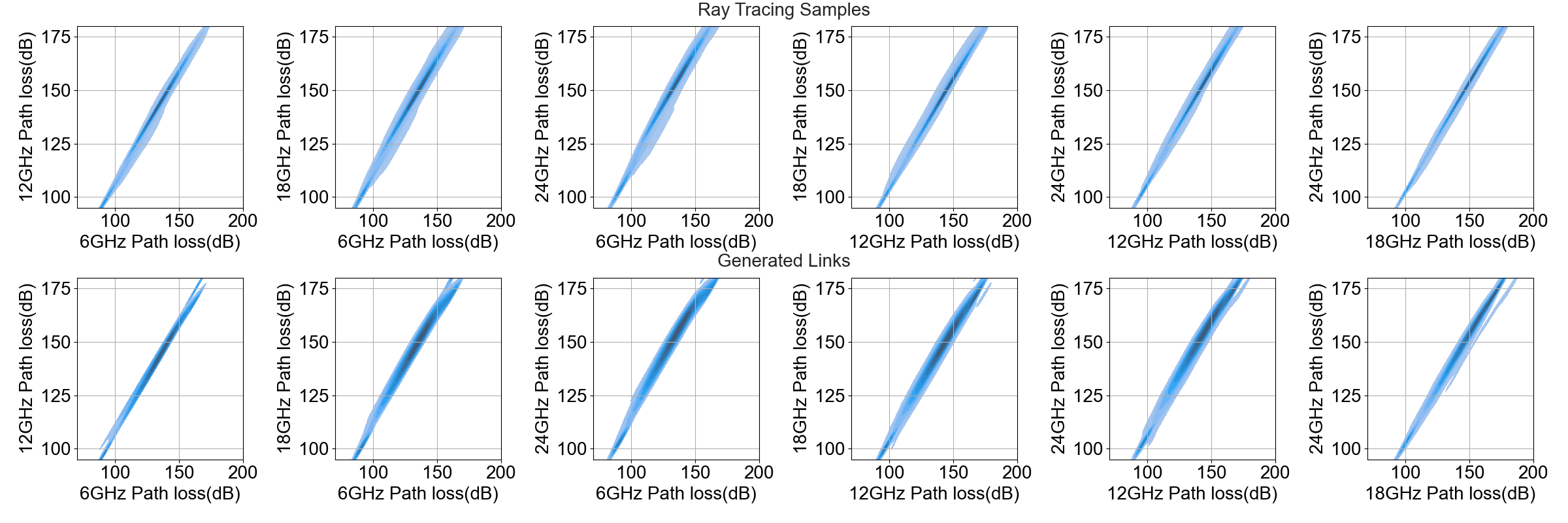}
  \caption{The KDE plot of bivariate distributions of the path loss at \SI{6}{GHz}, \SI{12}{GHz}, \SI{18}{GHz}, and \SI{24}{GHz}.}
  \label{fig:received_power_comparing}
\end{figure*}

Our initial assessment focuses on the model's capability to accurately reflect the joint distribution of path loss across two specific frequencies.
The evaluation is conducted as follows:
Let $\tilde{\boldsymbol{x}}=\nbG(\boldsymbol{c},\boldsymbol{z})$ represent the generated model output, and let $(\nbc_i,\nbx_i)$, $i=1,\ldots,N_{\rm ts}$ represent the test samples, where each data point consists of the link condition $\nbc_i$ and the corresponding path data vector $\nbx_i$. 
For each test sample, we evaluate the vector
\begin{equation} \label{eq:vdata}
    \nbv_i = (v_{i1},v_{i2},v_{i3},v_{i4}) = (\phi_1(\nbx_1), \phi_2(\nbx_2),\phi_3(\nbx_3), \phi_4(\nbx_4)),
\end{equation}
where $v_{ij}$ denotes the omni-directional path loss for sample $i$ at frequency $j$ for $j=1,2,3,4$.
Here, $\phi_j(\nbx)$ calculates the omni-directional path loss at frequency $j$ from the path vector $\nbx$.
The omni-directional path loss reflects the path loss that would occur if both the gNB and UE possessed omni-directional antennas.

To assess these values against those generated by the model, for each test sample $i$, a random sample $\nbx_i^{\text{rnd}} = \nbG(\nbc_i,\nbz_i)$ is generated by the trained generator $\nbG$ using a random $\nbz_i$ under the same conditions $\nbc_i$ as the test data.
Then, we compute the set of generated path losses
\begin{align}
    \label{eq:vgen}
    \nbv_i^{\rm rnd} = & (v_{i1}^{\rm rnd},v_{i2}^{\rm rnd},v_{i3}^{\rm rnd},v_{i4}^{\rm rnd}) \nonumber \\
    = & (\phi_1(\nbx_1^{\rm rnd}), \phi_2(\nbx_2^{\rm rnd}), \phi_3(\nbx_3^{\rm rnd}), \phi_4(\nbx_4^{\rm rnd})).
\end{align}
The distribution of $\nbv_i$ and $\nbv_i^{\rm rnd}$ should be closely matched.

Fig.~\ref{fig:received_power_cdf} presents the empirical cumulative distribution functions (CDFs) for the marginal distributions of both ray tracing ($v_{ij}$) and model-generated data $v_{ij}^{\rm rnd}$ across the four frequencies, i.e., 6, 12, 18, and 24 GHz. As can be seen from the plot, the marginal CDFs show great much at all considered frequencies.

Furthermore, a notable observation is that the generator also effectively captures the joint statistics. 
The first row of Fig.~\ref{fig:received_power_comparing} depicts kernel density estimation (KDE) plots of the test data points $(v_{i1},v_{i2},v_{i3},v_{i4})$ at different frequencies.
And the second row shows KDE plots of the generated outputs $(v_{i1}^{\rm rnd},v_{i2}^{\rm rnd}, v_{i3}^{\rm rnd},v_{i4}^{\rm rnd})$. 
From the KDE results, we can observe that the \emph{joint} distributions are also well captured by the proposed scheme.

\subsection{SNR with Beamforming}
\begin{figure}
  \centering
  \includegraphics[width=1 \linewidth]{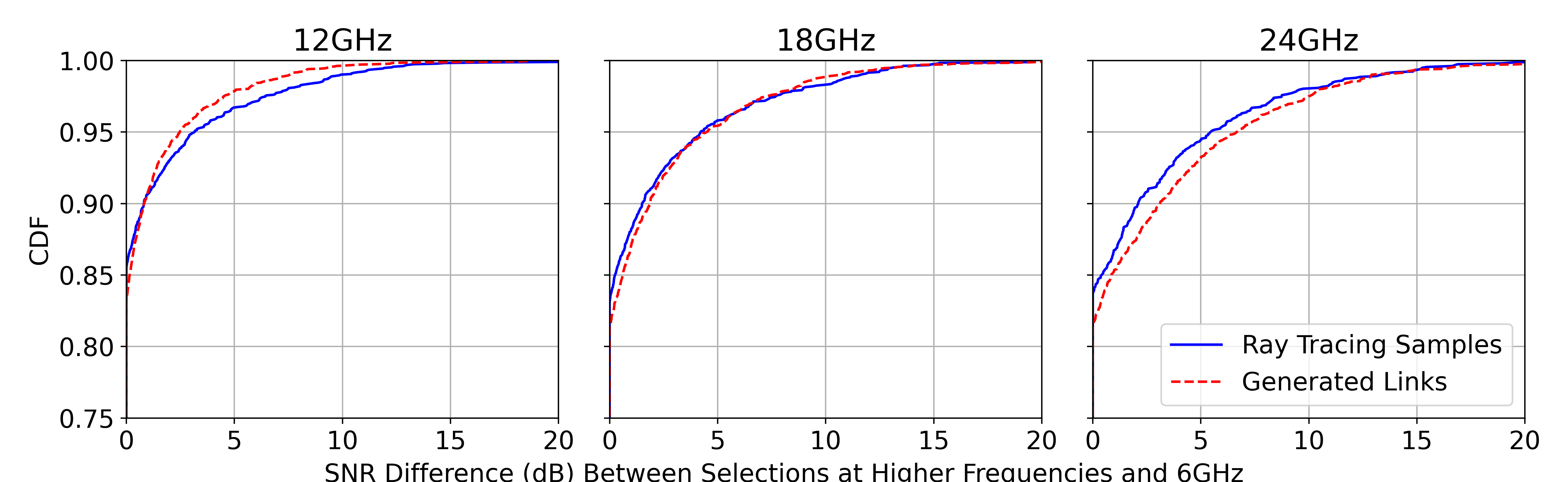}
  \caption{The CDF plot of SNR differences (in dB) between beam selected by higher frequencies (\SI{12}{GHz}, \SI{18}{GHz}, and \SI{24}{GHz}) and beam selected by \SI{6}{GHz}.}
  \label{fig:snr-beam-selected}
  \vspace{-0.45cm}
\end{figure}

\begin{table}[t]
\caption{Multi-frequency SNR Simulation Parameters}
\label{tab:snr_simu_param}
\small 
\setlength{\tabcolsep}{3pt} 
\renewcommand{\arraystretch}{1.0} 
\begin{tabularx}{\columnwidth}{@{}lXXXX@{}}
\toprule
\textbf{Parameter} & \textbf{6 GHz} & \textbf{12 GHz} & \textbf{18 GHz} & \textbf{24 GHz} \\
\midrule
Bandwidth [MHz] & 100 & 200 & 300 & 400 \\
gNB URA dims. & 2 $\times$ 2 & 4 $\times$ 4 & 5 $\times$ 5 & 7 $\times$ 7 \\
UE ULA dims. & 1 $\times$ 2 & 1 $\times$ 2 & 1 $\times$ 3 & 1 $\times$ 3 \\
\addlinespace 
Max. TX Power (dBm) & \multicolumn{4}{c}{33 (3GPP TR 38.141 \cite{3gpp38141})} \\
Sectors for gNBs & \multicolumn{4}{c}{3} \\
UE Noise Figure (dB) & \multicolumn{4}{c}{7} \\
Antenna Pattern & \multicolumn{4}{c}{3GPP TR 37.840 \cite{3gpp37840}} \\
Down-tilt Angle & \multicolumn{4}{c}{12°} \\
\bottomrule
\end{tabularx}
\vspace{-0.5cm}
\end{table}


To further evaluate the effectiveness of the proposed method, we estimate the AoA and AoD using a reliable lower frequency, specifically \SI{6}{GHz}, and subsequently apply the derived beamforming vector to higher frequencies: \SI{12}{GHz}, \SI{18}{GHz}, and \SI{24}{GHz}. We summarize the simulation parameters for prospective cellular systems at these frequencies in Table~\ref{tab:snr_simu_param}. The bandwidth allocation for each frequency scales proportionally with the carrier frequency, in line with current deployment strategies. We then calculate the signal-to-noise ratio (SNR) difference between the links generated and those obtained via ray-tracing samples. Figure~\ref{fig:snr-beam-selected} illustrates that the CDF of the SNR difference between the test and generated data aligns closely across different frequencies.

\subsection{Root Mean Square (RMS) Angles Spread}
\begin{figure}
  \centering
  \includegraphics[width=.99 \linewidth]{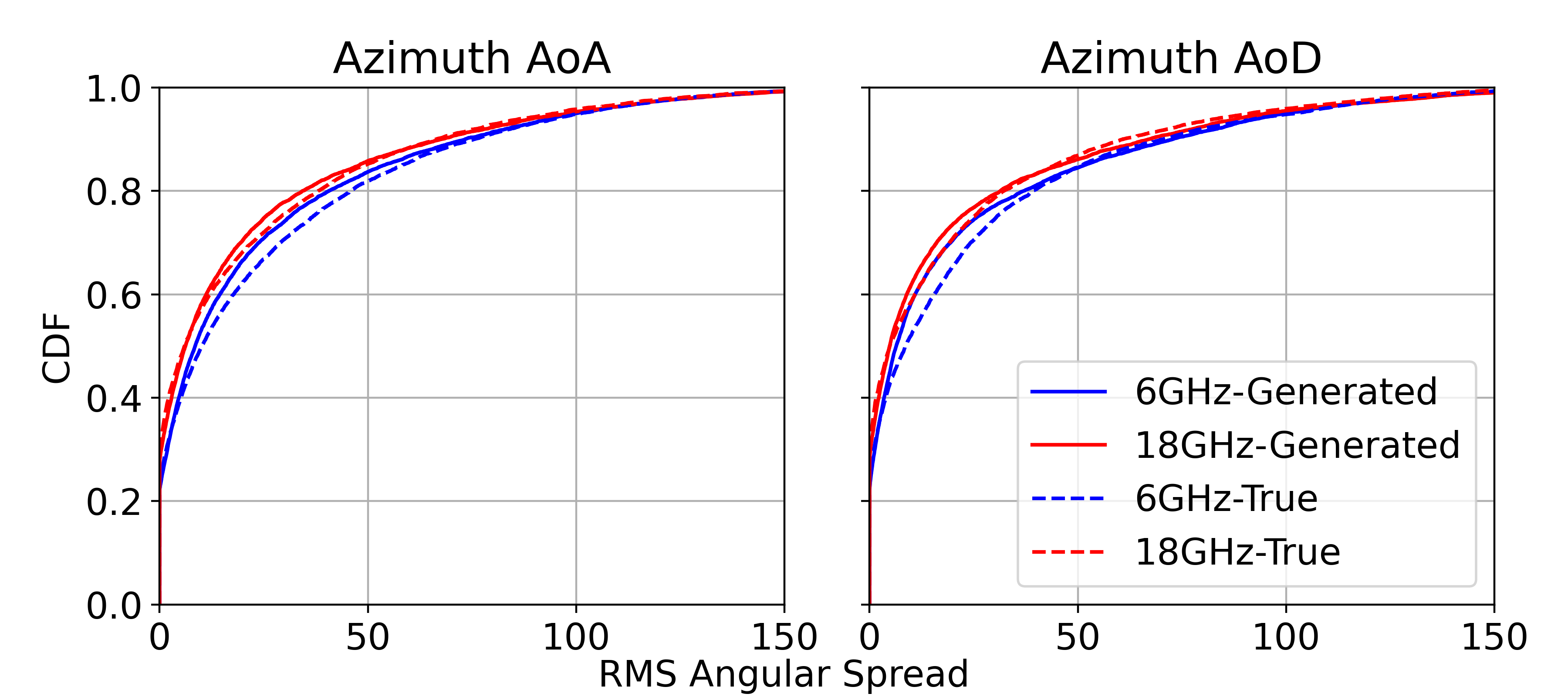}
  \caption{RMS azimuth AoA/AoD spread, including the zero values}
  \label{fig:rms_angular}
  \vspace{-0.5cm}
\end{figure}

The Root Mean Square (RMS) angle spread is critical for evaluating channel models, particularly in the FR3 bands. Figure~\ref{fig:rms_angular} displays CDF images of the RMS spreads for AoA and AoD. Moreover, to address heterogeneous data types, we align path angles to the LOS direction. 
Specifically, for preprocessing the AoA, the LOS AoA direction serves as the z-axis in a new spherical coordinate system, facilitating the calculation of azimuth and inclination transformation angles. A similar approach is adopted for the AoD, using the LOS AoD direction. This alignment enables the network to more effectively learn a statistical model relative to the condition vector. 
According to Figure~\ref{fig:rms_angular}, the model successfully captures the RMS angle spread characteristics at both carrier frequencies, with the spread at \SI{6}{GHz} being slightly greater than at \SI{18}{GHz}.

\section{Conclusion}
\label{sec:conclusion}


This work proposed a GAN-based channel modeling scheme to capture the multi-frequency characteristics at upper mid-band/FR3. With the link state predictor stage, the proposed generative neural network can effectively reflect both marginal and joint channel distributions. The agreements of path loss CDF, KDF, beamforming SNR, as well as RMS distribution together showed great potential for the use of GAN in FR3 channel modeling.


\bibliographystyle{IEEEtran}
\bibliography{bibl}

\end{document}